\documentclass[prl,twocolumn,superscriptaddress]{revtex4}

\usepackage{epsfig}

\newcommand{\udar}{\updownarrow}
\newcommand{\lrar}{\leftrightarrow}
\newcommand{\ran}{\rangle}

\begin{document}

\title{Methods for scalable optical quantum computation }

\author{Tal Mor}
\affiliation{Computer Science Department, Technion, Haifa 32000,
Israel}
\author{Nadav Yoran}
\affiliation{Computer Science Department, Technion, Haifa 32000,
Israel} \affiliation{H.H.Wills Physics Laboratory, University of
Bristol, Tyndall Avenue, Bristol BS8 1TL, UK}
\date{\today}






\begin{abstract}
We propose a scalable method for implementing linear optics quantum
computation using the ``linked-state'' approach. Our method avoids the
two-dimensional spread of errors occurring in the preparation of the
linked-state. Consequently, a proof is given for the scalability
of this modified linked-state model, and an exact expression for
the efficiency of the method is obtained.
Moreover, a considerable improvement in the efficiency,
relative to the original linked-state method, is achieved.
The proposed method is applicable to  Nielsen's
optical ``cluster-state'' approach as well.
\end{abstract}

\maketitle

Although very successful in implementing various protocols of
quantum information, optical approaches
were not considered to be a useful tool for implementing scalable
quantum computation. The reason is the lack of interaction between
photons required for the implementation of two-qubit gates. It
was, however, shown by Knill, Laflamme and Milburn (KLM)
\cite{KLM} that this obstacle can be overcome. KLM have shown that
a probabilistic two-qubit gate can be implemented by means of
interference with ancillary photons and incomplete measurements
(detecting only a part of the photons). The resulting {\em
conclusive measurement} or {\em unambiguous detection} sometimes
yields the correct quantum output, and the user can identify these
instances as these are accompanied by a specific classical outcome
of the measurement. The probability of success of any KLM-type gate
is determined by the complexity of the ancillary state. The more
resources, in terms of the number of elementary operations,
invested in the preparation of the ancilla, the higher the success
probability. The basic requirements of the KLM scheme are
single-photon sources, photon counters (that distinguish between
different number states) and fast feed-forward ability. It was
shown by KLM that by using their gates scalable quantum
computation can be implemented. Yet, a huge resource overhead is
required in order to reduce the unavoidable probability to fail in
each gate application. Additionally, scalability is achieved (even
in the ideal case where all optical devices are error-free) only
if error correcting codes are employed.

The problem of the unavoidable possibility of failing occurring in
each KLM-type gate (which leads to the enormous resource overhead) was
recently solved by a new scheme~\cite{YR03}. The
information processing in this scheme is done in two stages, which
we call here ``construction'' and ``evolution''. In the
construction stage a multi-photon ``linked-state'' is prepared,
corresponding to the specific computation. The subsequent
evolution stage is then fully deterministic.
The linked-state consists of
chains of photons, each chain corresponding to a single logical
qubit. In such a
chain the path degree of freedom of each photon is maximally
entangled with the polarization of the next photon. The
construction stage includes the preparation of the above chains and
the ``weaving'' of the chains together into the overall linked-state.
The weaving steps are done
by entangling photons of different chains according to the circuit
one wishes to process. Both the chain-preparation and the
weaving are done in parallel using KLM-type gates via
unambiguous detection. In the evolution stage the data, which is
initially encoded in the polarization of the first photon in the
chain, progresses along the chain (from the polarization of one
photon to that of the next photon) by a sequence of teleportation
steps. In each such teleportation
step both (path and polarization)
degrees of freedom of a single photon are
measured in the Bell basis; a single teleportation of this type
was experimentally demonstrated in~\cite{Demar}.
One-qubit gates are implemented by rotating the
polarization of the photon that carries the data prior to
teleporting it.
A conditional phase gate (CPHASE) is induced on two logical
qubits in any event where the data in the two corresponding chains
passes through two photons that were entangled by a weaving step.

The linked-state can therefore be viewed as a two dimensional
structure consisting of connected chains. This method exhibits
vastly improved efficiency over the KLM method (see~\cite{YR03}),
and furthermore, it is believed that this scheme is scalable even
without the need for inherent error correction. Yet, so far, due
to the complex dynamics of the construction process there is no
proof for the scalability of the linked-state scheme, let alone a
calculation of its efficiency. When a KLM-type gate fails and a
photon is removed, the attempt to replace it by a new photon may
also fail, resulting in the removal of one of the photons that
were connected to it as well. A failure can, therefore, spread in
this manner both backwards in one chain and, worse, also to other
chains. Efficiency results appearing in \cite{YR03} were based on
simulations of a simple circuit where two-qubit gates are
repeatedly applied to just a pair of logical qubits. A full proof
of scalability requires the analysis of a rather complex
2-dimensional random walk process, and was not done yet.

A different yet somewhat similar approach, suggested by
Nielsen~\cite{Niel}, is the optical version of the cluster-state
model of Raussendorf and Briegel \cite{RB} (further developments
are described in~\cite{btr}).
This method also requires the preparation of a
multi-photon 2-dimensional structure of inter-connected chains.
It should be noted that a previous estimation~\cite{Niel} of the
efficiency of his method does not take into account all
(two-dimesnional) paths through which a gate failure can spread
--- measured photons that have more than a single connection to
the rest of the cluster are more expansive to replace.

Here we present a {\em scalable linked-state method} and prove its
scalability. The key feature of the method is that it avoids the
two-dimensional spread of errors. Therefore, it avoids the need
for a complex 2-dimensional random walk analysis. In our method,
failures in probabilistic gates that might spread backwards are
confined to the chain preparation steps.
The connections between
the chains are established in a manner that prevents any spread of
failures during the weaving steps. Consequently, we are able to
provide an exact expression for the efficiency of the method. In
addition, the new linked-state method is considerably more
efficient then the original one.
The proposed method can be applied, with the same advantages,
to design a {\em scalable optical cluster-state method} as well.

The basic idea is to construct for each logical qubit a somewhat
more complex chain of photons (than in~\cite{YR03}). The
polarization of each such ``linked-photon'' in such a chain is
entangled not only with the path degree of freedom of the preceding
linked-photon in the chain, but also with the path of an additional
photon, a ``free-arm'' photon.
A schematic description of the chain is given
in Fig. 1.
Only the free-arm photons are then used in the weaving steps
to establish a connection between two chains. This is done by
applying to these photons a probabilistic KLM-type gate. A
successful application of the gate will produce the entangled
state required for the application of a two-qubit CPHASE gate as
in the original linked-state method. A failure of a KLM-type gate
results in the measurement of (exactly) one of its input photons. Such a
failure in the connecting operation will therefore result in the
measurement
of one of the free-arm photons.
However, no other part of that
chain will be affected. Therefore the transmission of the data
along the chain is still possible, and a second attempt to
establish a connection can be applied to the next free-arm photon
in the chain, and so on.
The state of such a chain is the following
\begin{eqnarray}\label{1-chain}
 &|\udar\ran_{p_1}
 \biggl(|0\ran_{p_{1}}|\udar\ran_{p_{2}}|+\ran_{p_{2^{\prime}}}+\,
 |1\ran_{p_{1}}|\lrar\ran_{p_{2}}|-\ran_{p_{2^{\prime}}}\biggr)
 \nonumber \\
 &\biggl(|0\ran_{p_{2}}|\udar\ran_{p_{3}}|+\ran_{p_{3^{\prime}}}+\,
 |1\ran_{p_{2}}|\lrar\ran_{p_{3}}|-\ran_{p_{3^{\prime}}}\biggr)
 \cdots \\
 &\biggl(|0\ran_{p_{n-1}}|\udar\ran_{p_{n}}
 |+\ran_{p_{n^{\prime}}}+\,|1\ran_{p_{n-1}}|\lrar\ran_{p_{n}}
 |-\ran_{p_{n^{\prime}}}\biggr)|0\ran_{p_{n}} \nonumber
\end{eqnarray}
where $\{ |0\ran\,,\, |1\ran \}$ and $\{ |\udar\ran\,,\,|\lrar\ran
\}$ denote the path and polarization of each photon respectively;
and $|+\ran = (|0\ran+|1\ran)/\sqrt{2}$ and $|-\ran =
(|0\ran-|1\ran)/\sqrt{2}$ are the states of the photons that
constitute the free-arms (denoted by primed indices). The terms
within the brackets are the links of the chain, carried by three
physical degrees of freedom
(of three different photons) in a GHZ state.

\begin{figure}\begin{center}
\epsfig{file=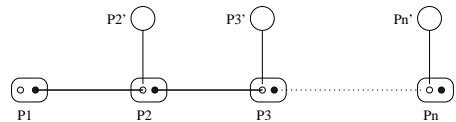}
\caption{\small  A chain state for our linked state model.
         The rectangles represent the linked-photons whose polarization
         and path degree of freedom are denoted by empty and full
         circles respectively. The connected circles
         (denoted with primes)
         signify the free-arm photons of which only the path degree of
         freedom is utilized. }
\end{center}\end{figure}

Let us now discuss in detail a single weaving step connecting two
chains using the free-arm photons. The initial states of the links
that are to be connected are given by:
$(|0\ran_{p_{1}}|\udar\ran_{p_{2}}|+\ran_{p_{2^{\prime}}}
 +|1\ran_{p_{1}}|\lrar\ran_{p_{2}}|-\ran_{p_{2^{\prime}}})$ and
$(|0\ran_{q_{1}}|\udar\ran_{q_{2}}|+\ran_{q_{2^{\prime}}}
 +|1\ran_{q_{1}}|\lrar\ran_{q_{2}}|-\ran_{q_{2^{\prime}}})$.
The desired final state -- after measuring $p^{\prime}$ and $q^{\prime}$
-- is the following:
\begin{eqnarray}
 &|0_{p_{1}}\udar_{p_{2}}\ran \:
 (\,|0_{q_{1}}\udar_{q_{2}}\ran\, + |1_{q_{1}}\lrar_{q_{2}}\ran\,) & +
 \nonumber \\
 &|1_{p_{1}}\lrar_{p_{2}}\ran \:
 (\,|0_{q_{1}}\udar_{q_{2}}\ran\, - |1_{q_{1}}\lrar_{q_{2}}\ran\,) &
 \label{1.5}
\end{eqnarray}
This entangled state of four degrees of freedom of four different
photons is the basic unit which is required for the application of
a two-qubit gate in the linked-state model \cite{YR03}. This
entanglement between the links is established by applying a
KLM-type CPHASE gate to $p_{2^{\prime}}$ and $q_{2^{\prime}}$.
This gate~\cite{KLM} consists of two teleportation protocols based
on the Fourier transform ($\hat{F}_{n+1}$). In addition to one
input photon, $n$ ancillary photons take part in each of these
``$F$-teleportation'' protocols
which succeed or fail independently -- the success probability of
each is $n/(n+1)$.
The KLM-type gate succeeds if the two $F$-teleportations succeed,
thus the success probability of this (CPHASE) gate is
$n^2/(n+1)^2$. We shall henceforth denote such gates as
$CZ_{(n)}$.
The conditional gate operation is achieved due to
the entangled state ($|CS_{n}\ran$) of the $2n$ ancillary photons.
As defined in \cite{KLM}, $|CS_{n}\ran=
\sum_{i,j=0}^{n}(-1)^{(n-i)(n-j)}|t_{n}^{i}\ran
 |t_{n}^{j}\ran$,
where $
|t_{n}^{i}\ran=|1\ran^{i}|0\ran^{n-i}|0\ran^{i}|1\ran^{n-i}$ is
the state of $2n$ modes carrying exactly $n$ photons, that take
part in one $F$-teleportation (in $|t_{n}^i\ran$ unlike the rest
of the paper, $\{|0\ran,|1\ran\}$ are states of zero and one
photon in a mode). If the operation of $CZ_{(n)}$
succeeds
then a negative phase is introduced to the
state $|1\ran_{p_{2}^{\prime}}|1\ran_{q_{2}^{\prime}}$,
and the state of the overall system becomes:
\begin{eqnarray}
 |0\udar\ran_{p_{1}p_{2}}|0\udar\ran_{q_{1}q_{2}}
 \biggl( |00\ran + |01\ran
   +|10\ran - |11\ran \biggr)_{p_{2^{\prime}}q_{2^{\prime}}}
 + \nonumber \\
 |0\udar\ran_{p_{1}p_{2}}|1\lrar\ran_{q_{1}q_{2}}
  \biggl( |00\ran - |01\ran
    +|10\ran + |11\ran \biggr)_{p_{2^{\prime}}q_{2^{\prime}}}
 + \nonumber \\
 |1\lrar\ran_{p_{1}p_{2}}|0\udar\ran_{q_{1}q_{2}}
 \biggl( |00\ran + |01\ran
        -|10\ran + |11\ran \biggr)_{p_{2^{\prime}}q_{2^{\prime}}}
 + \nonumber \\
 |1\lrar\ran_{p_{1}p_{2}}|1\lrar\ran_{q_{1}q_{2}}
 \biggl( |00\ran - |01\ran
        -|10\ran - |11\ran \biggr)_{p_{2^{\prime}}q_{2^{\prime}}}
\label{2}
\end{eqnarray}
At this stage photons $p_{2^{\prime}}$ and $q_{2^{\prime}}$ are
measured in the basis $\{|+\ran,\,|-\ran\}$ (the $x$-basis). Taking
for example the first line in (\ref{2}), the term within the brackets
can be written as $(|++\ran\,+\,|+-\ran\,+\,|-+\ran\,-\,|--\ran)$.
The terms within the brackets in the other lines correspond to similar
expressions only with the minus sign appearing in different places.
Therefore, upon receiving the outcome $|++\ran$
in the measurement, the resulting state of photons
$\{p_{1},p_{2}\}$ and $\{q_{1},q_{2}\}$ would be exactly as
in Eq. (\ref{1.5}),
while other results would lead to similar states up to local phase
adjustments. A failure in the operation of $CZ_{(n)}$
would
result in the measurement in the $z$-basis of either
$p_{2^{\prime}}$ or $q_{2^{\prime}}$.
Taking $p_{2^{\prime}}$ for example, we can see
from the initial state of the link that whether we get an outcome
of $|0\ran$ or $|1\ran$, the path of $p_{1}$ will remain maximally
entangled with the polarization of $p_{2}$, enabling the
transmission of the data forward to the next link.

Let us address the construction of the chains.
The basic units from which the chain is constructed are
``two-photon units'' consisting of a linked-photon ($p_i$) and a
free-arm photon ($p_i'$) in a maximally entangled state.
Such two-photon units
can be produced from four single (non-entangled) photons, by using
for example the gate $\mbox {c-z}_{1/16}$ of KLM~\cite{KLM}.
Thus, the basic requirements of our
scheme are similar to those of the KLM-scheme.
In each step in the construction an attempt is made to connect
such a two-photon unit to the already constructed chain (Fig.~1).
One can entangle, by using CPHASE KLM-type gates ---
$CZ_{(n)}$,
the path degree of freedom of the last photon in the chain
to the polarization of the linked-photon in the new unit.
Since the polarization
of the last photon in the chain is already entangled with the path
of the previous photon, and this entanglement must be kept, two
$CZ_{(n)}$ gates
must be applied -- one for each of the
polarization modes
of the photon \cite{YR03}.
The operation, therefore, includes four
$F$-teleportations. Clearly, it is
more efficient
to apply first the two teleportations on the new pair,
minimizing the risk of removing photons from the chain.
Thus, for each random walk step we have an ``off-line'' preparation
in which (the first) two $F$-teleportation steps
are applied to a two-photon unit
using two $|CS_{n}\rangle$ states. The resulting ``prepared unit''
will then be used when applying the two $F$-teleportation to the last photon
in the chain.

The efficiency of the new method can be readily calculated since
the construction of a chain is simply a one dimensional random
walk process. A step forward is taken when the operation of
connecting a new two-photon unit to the chain has succeeded. A
step backwards is taken when this operation fails in such a way
that the last linked-photon in the chain was measured (removing
also a free-arm photon).
Because of the fact that a separate $F$-teleportation is applied
to each of the polarizations of the last photon in the chain,
there is also a possibility to fail {\em without} measuring the
last photon in the chain, leaving the chain unaffected (while
destroying just the prepared-unit). In the random walk analysis
only the two $F$-teleportations applied to the last photon in the
chain are considered.
Therefore the probability of
success is $p=n^{2}/(n+1)^{2}$.
The probability to fail destructively and remove a linked photon is
$q=(1-p)/2=(2n+1)/2(n+1)^{2}$ (see~\cite{YR03}).
For a one dimensional
random walk the average number of attempts required in order to
advance one step forward (for a large number of steps) is simply
$R(n)=1/(p-q)$.
In terms of $n$ --- the parameter of the KLM-type
gate --- we obtain
$R(n)=2(n+1)^{2}/(2n^{2}-2n-1)$. In order to obtain the amount of
resources, in terms of ancillary states, consumed in the process,
we need to calculate the cost of a single random walk step. Since
one applies two $F$-teleportations on the two-photon
unit, before a similar operation is applied on the last photon in
the chain, on average we will need $(n+1)^{2}/n^{2}$ attempts in
order to succeed, each costing a single two-photon unit. In each
of these attempts one $|CS_{n}\ran$ (in case the first
$F$-teleportation fails) or two such states (in all other cases) are
consumed as well. On average in each random walk step -- whether
successful or not -- $(n+1)^{2}/n^{2}$ two-photon units and
$(2n+1)(n+1)/n^{2}$ copies of $|CS_{n}\ran$ are consumed.
The resources required on average in order to add a two photon
unit to the chain are given by multiplying the above expressions
by $R(n)$:
\begin{eqnarray}
 &\frac{2(n+1)^{4}}{n^{2}(2n^{2}-2n-1)} \quad
 \mbox{$2$-photon units} \nonumber \\
&\frac{2(n+1)^{3}(2n+1)}{n^{2}(2n^{2}-2n-1)} \,\,
 \mbox{copies of} \,\,|CS_{n}\ran
\label{4}
\end{eqnarray}

What is the average number of links, per two-qubit gate, that a
chain must include? As each of the independent $F$-teleportations
constituting a $CZ_{(m)}$ gate succeeds with probability of
$m/(m+1)$, for each two-qubit gate a chain must have, on average,
$(m+1)/m$ links with free arms.
We can now summarize the efficiency analysis.
On average, in order to implement
a single two-qubit gate --- that is to construct the required
length in two chains and connect them --- the following resources
are required (for a large number of gates):
\begin{eqnarray}
 &\frac{2(m+1)}{m}\frac{2(n+1)^{3}(2n+1)}{n^{2}
 (2n^{2}-2n-1)} \;\; \mbox{copies of}\;\; |CS_{n}\ran   \nonumber \\
 &\frac{2(m+1)}{m}\frac{2(n+1)^{4}}{n^{2}(2n^{2}-2n-1)}
 \;\; \mbox{$2$-photon units} \nonumber \\
 &\frac{(m+1)^{2}}{m^{2}} \;\; \mbox{copies of}\;\;\; |CS_{m}\ran
 \label{4.5}
\end{eqnarray}
where chain construction is based on $CZ_{(n)}$ gates
and the connections between the chains are carried out
using $CZ_{(m)}$ gates.

\begin{figure}\begin{center}
\epsfig{file=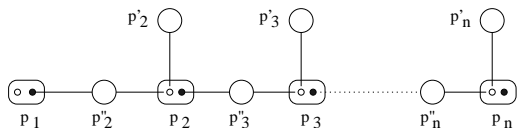}
\caption{\small
 A chain state for the linked-state model with inert
         photons (denoted by double primes)
         connected in between the linked-photons.}
\end{center}\end{figure}

As can be easily verified, in the random walk process we obtain
$p>q$ when using any
$CZ_{(n)}$
with $n\geq 2$, which means that the simplest ancillary states
with which a chain can be efficiently prepared are the $4$-photon
states $|CS_{2}\ran$. (For comparison, in the original
method~\cite{YR03} the simplest KLM-type gate, with which the
linked-state can be constructed in a straightforward way, requires
the state $|CS_{3}\ran$.) The process of establishing a connection
between two chains can use any KLM-type gate  (namely, $CZ_{(m)}$
with $m\geq 1$),
as it does not affect any part of the chains except the free
arms.

In order to improve the efficiency of our method
one can employ ``inert photons'' as was done in the original
scheme \cite{YR03}. These are additional photons included in the
chain, that do not take part in the operation of the logic gates.
Their only purpose is to improve the bias of the random walk
process, by decreasing the risk of losing the last linked-photon
in the chain.
Constructing the chain in this way one adds in each step a
$3$-photon unit (as shown in Fig. 2) --- the linked-photon with
its free-arm and an inert photon. Clearly, each random walk step
would now include more operations, however, at least for low
values of $n$ the overall construction would be more efficient.
Alternatively, one can improve the random walk bias of the basic
free-arm method by adding to the chain a number of connected
two-photon units in each step. By using each of these methods it
is possible to construct the chain using the simplest KLM-type gate,
$CZ_{(1)}$~\cite{MY-future}.

The ``free-arm photon'' method can be applied to the optical
cluster state model as well. In this model each photon carries a
single relevant two-dimensional degree of freedom and a chain is
constructed by entangling each photon to its two nearest
neighbors. As in the linked-state model, here also the data is
being processed while progressing along the chain. However this is
done not through measurements in the Bell-basis but by
measurements of single photons (i.e. a single degree
of freedom). Due to the different structure of the chain
single-qubit gates are not implemented directly
but through single-photon measurements as
well.

\begin{figure}\begin{center}
\epsfig{file=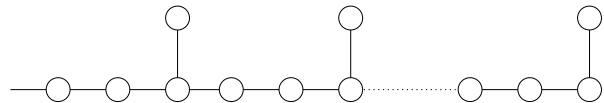} \caption{\small A chain state for the
cluster state model.}
\end{center}\end{figure}

We suggest now a scalable ``free-arm optical cluster state
model''. The chain that we consider is shown in Fig.~3. Each third
photon is connected to free-arm photon. After connecting it to
another free-arm photon of a different chain the resulting
four-photon state can be used to implement a conditional phase
flip gate. The photons without free-arms are required in order to
apply a (general) single-qubit gate, between any two consecutive
two-qubit gates. A simple way to construct the above chain is, for
example, to add in each step a $4$-photon unit as depicted in Fig.
3. If an attempt to add such a unit fails then one can still try
and fix the chain (damaged because of this failure). Only a series
of three gate failures would destroy the last $4$-photon unit in
the chain.
The resources required for the addition of one $4$-photon unit to the
chain are $(n+1)^{4}/(n^{2}(n+1)^{2}-n)$ copies of $4$-photon units
and $(n+1)^{2}(n^{2}+3n+3)/(n^{2}(n+1)^{2}-n)$ copies of $|CS_n\ran$,
with $n\geq 1$ (details are given in~\cite{MY-future}).

The scalable models that we have presented above
can be used to implement general-purpose quantum computation:
instead of preparing the custom-made linked (or cluster)
state which corresponds to the specific computation, it would be
sufficient to construct standard-form chains. The weaving of the
chains can then be carried out in conjunction with
the evolution stage. Each connection between chains can be
established just before the corresponding two-qubit gate is
executed (that is, before the connected photons are measured). As
the data carried by the chains is not affected by a failed
connection attempt, one has simply to re-try to connect the chains
using the next free-arm. Clearly, in this way, given that a long
enough chain has been prepared for each qubit, any computation can
be performed with the same efficiency as calculated above.

\acknowledgments     

This work was supported in parts by the Israel Science Foundation
-- FIRST (grant \#4088103). N.~Y. acknowledges the Aly Kaufman
Fellowship, Technion, and the UK EPSRC (grant GR/527405/01).



\end{document}